\documentclass[journal,10pt]{IEEEtran}
\ifCLASSINFOpdf
\else
   \usepackage[dvips]{graphicx}
\fi
\usepackage{url}

\usepackage{enumitem}
\usepackage{array}
\usepackage{booktabs}
\usepackage{subfigure}
\usepackage{pdfpages}
\usepackage{multirow}
\usepackage{amssymb}
\usepackage{tablefootnote}
\usepackage{bbm}
\usepackage{algorithm, algorithmic}
\usepackage{graphicx}
\usepackage{cite}
\usepackage{amsmath}
\begin{document}
\title{

Swin-BERT: A Feature Fusion System designed for Speech-based Alzheimer's Dementia Detection}

\author{Yilin Pan,
        Yanpei Shi*,
        Yijia Zhang,
        Mingyu Lu
\thanks{Yilin~Pan (email: yilin.pan@dlmu.edu.cn) is with the College of Artificial Intelligence, Dalian Maritime University, China; 
Yanpei~Shi* (corresponding author, shiyanpei0826@gmail.com) is with the Department of Computer Science, University of Sheffield, United Kingdom;
Yijia~Zhang is with College of Information Science and Technology,  Dalian Maritime University, China;
Mingyu~Lu is with the College of Artificial Intelligence, Dalian Maritime University, China.}

}
\markboth{Journal of \LaTeX\ Class Files, July 2024}
{Shell \MakeLowercase{\textit{et al.}}: Bare Demo of IEEEtran.cls for IEEE Journals}
\maketitle

\begin{abstract}
Speech is usually used for constructing an automatic Alzheimer's dementia (AD) detection system, as the acoustic and linguistic abilities show a decline in people living with AD at the early stages. However, speech includes not only AD-related local and global information but also other information unrelated to cognitive status, such as age and gender. In this paper, we propose a speech-based system named Swin-BERT for automatic dementia detection. For the acoustic part, the shifted windows multi-head attention that proposed to extract local and global information from images, is used for designing our acoustic-based system. To decouple the effect of age and gender on acoustic feature extraction, they are used as an extra input of the designed acoustic system. For the linguistic part, the rhythm-related information, which varies significantly between people living with and without AD, is removed while transcribing the audio recordings into transcripts. To compensate for the removed rhythm-related information, the character-level transcripts are proposed to be used as the extra input of a word-level BERT-style system. Finally, the Swin-BERT combines the acoustic features learned from our proposed acoustic-based system with our linguistic-based system. The experiments are based on the two datasets provided by the international dementia detection challenges: the ADReSS and ADReSSo. The results show that both the proposed acoustic and linguistic systems can be better or comparable with previous research on the two datasets. Superior results are achieved by the proposed Swin-BERT system on the ADReSS and ADReSSo datasets, which are 85.58\% F-score and 87.32\% F-score respectively.

\end{abstract}
\section{Introduction}
\label{sec:introduction}

Automatic detection of Alzheimer's dementia (AD) from spontaneous speech has the potential to be applied as a cheap, easy-to-use, and objective clinical assistance tool for dementia detection. 
Speech-based automatic AD detection has shown its efficiency in enabling early treatment of people living with AD, which can slow down AD deterioration. Although the changes of a person\textquotesingle{s} speech can often be seen many years before diagnosis \cite{ross1990speech}, the audio recordings include multiple information, such as cognitive status, age \cite{duchesne2019effects,goy2016effects} and gender \cite{alsharhan2020investigating}. Lack of consideration of age and gender can affect the efficiency of the extracted acoustic features for dementia detection \cite{pan2021multitask,zaleta2010patient}. When learning the cognitive status related acoustic information in dementia detection, both the global and local acoustic related information have shown distinctive patterns throughout previous research clinically \cite{alzheimer20172017} or automatically \cite{pan2019automatic,chen2023speechformer++}. Swin Transformer \cite{liu2021swin} was designed to leverage the global and local information embedded in the image and speech \cite{wang2024speech} based shifted windows multi-head attention (SW-MHA), which is expected to be used for acoustic-based dementia detection in our paper.

As dementia progresses, almost all aspects of language can be affected \cite{groves2004comparison}. 
While constructing a linguistic-based system, the automatic word-level transcripts generated by an automatic speech recognition (ASR) system are generally used. In this process, the rhythm-related information, such as hesitations resulting from AD, is removed, although the rhythm-related information varies significantly between the healthy controls (HCs) and people living with AD \cite{meilan2020changes,pan2023ps,syed2020automated,yuan2020disfluencies,pan2021using}. A popular method for extracting rhythm-related information is by using the time alignment information estimated by the ASR system \cite{yuan2020disfluencies,pan2020improving}. As discussed in \cite{pan2020improving}, modelling the rhythm-related information as a sequential series can embed more information compared to the statistical analysis. In \cite{pan2023ps}, the sequential rhythm-related information is preserved by learning the embedding matrix of the character-level transcripts generated by an ASR system. In this paper, the sequential information embedded in the character-level transcripts and word-level transcripts is extracted with a BERT-style system for constructing a linguistic-based system.

Our contributions are as follows. 
\textbf{Firstly}, a SW-MHA inspired acoustic-based system is proposed to extract the global and local information related to dementia embedded within the audio recordings. Using the age and gender information as the extra input can improve the system's performance further. 
\textbf{Secondly}, using the character-level transcripts and word-level transcripts together is demonstrated to be informative for constructing a linguistic-based dementia detection system. 
\textbf{Finally}, a fusion system named Swin-BERT is designed to take advantage of the acoustic-based and linguistic-based systems to reach superior results on both the ADReSS datasets (released for the first international shared-task challenge \cite{LuzHaiderEtAl20ADReSS}) and ADReSSo (released for the second international shared-task challenge \cite{bib:LuzEtAl21ADReSSo}). 
\begin{figure*}[!htbp]
	\centering
	\includegraphics[width=\textwidth]{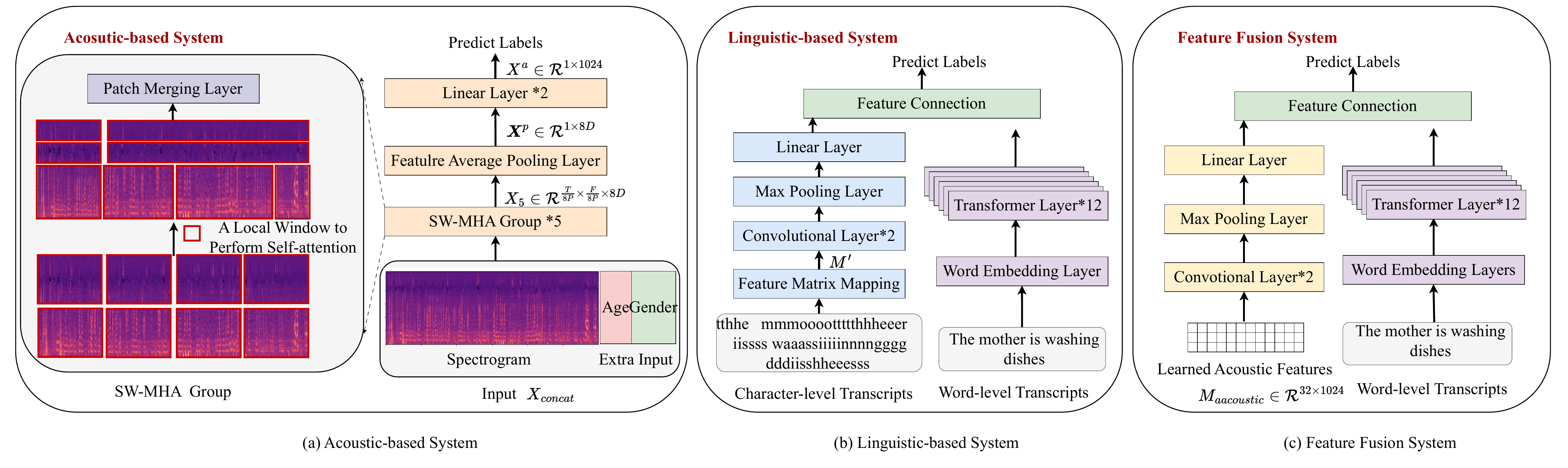}
	\caption{Designed linguistic-based, acoustic-based and feature fusion systems.}
	\label{fig:late_fusion}
\end{figure*}

\section{Designed Systems}
\label{systems}

This section describes the proposed systems for dementia detection, including the acoustic-based system, and linguistic-based system. The proposed feature fusion system named Swin-BERT is designed to leverage the extracted acoustic information and linguistic information, as is shown in Figure \ref{fig:late_fusion}. In this section, the structure of the acoustic-based and the linguistic-based systems are introduced in detail.

\subsection{Acoustic-based System}
\label{subsec:swin_transformer}

The acoustic-based system is composed of a hierarchical SW-MHA structure inspired by the Swin transformer, as shown in Figure \ref{fig:late_fusion} (a). Suppose $\boldsymbol X \in \mathcal {R}^{T \times F \times 1}$ denotes the input spectrogram of the model, where $T$ denotes the number of time frames, $F$ denotes the number of frequency bins, $1$ denotes the number of channels. To utilise the age and gender information, the age and gender information is first embedded into vectors, and the input spectrogram  $\boldsymbol X$ is concatenated with the provision of the age and gender embedding, and denoted as $\boldsymbol X_{concat}$. A SW-MHA block, together with a patch merging layer, is called a SW-MHA group. In our designed acoustic-based system, the processed spectrogram $\boldsymbol X_{concat}$ is processed by five SW-MHA groups for learning the global and local acoustic information related to dementia detection. The output of the first group is denoted as $\boldsymbol X^{1} \in \mathcal {R}^{\frac{T}{2P} \times \frac{F}{2P} \times 2D}$, and the output of the final SW-MHA group is denoted as $\boldsymbol X^{5} \in \mathcal {R}^{\frac{T}{8P} \times \frac{F}{8P} \times 8D}$. An average pooling operation is used on both time and frequency dimensions to output $\boldsymbol X^{p} \in \mathcal {R}^{1 \times 8D}$ before feeding into two linear layers for doing classification.

\subsection{Linguistic-based System}
\label{subsec:linguistic_only}
The designed linguistic-based system with word-level and character-level transcripts as the input of the BERT-based system is shown in Figure \ref{fig:late_fusion} (b). To extract the distinguishing information between the HCs and people living with AD from the word-level and character-level transcripts. First of all, the generated character-level transcripts are processed with a character dictionary to arrive at a character-level feature matrix. The character dictionary is composed of 32 characters. Each character-level transcript is transmitted into a one-hot matrix $M \in \mathcal {R}^{l \times 32}$, where $l$  equals to the character length. The length $l$ of each transcript depends on the input audio. To obtain a fixed length matrix $M'$, the character-level embedding matrix $M$ is chunked or padded with zero into the same length. Then, both the character-level embedding matrix and the word-level transcript have been used as the system’s input. To process the character-level embedding matrix, two convolutional layers – a max-pooling layer and a linear layer – are designed. One embedding layer and twelve transformer layers are used to extract word-level information from the transcripts. The processed character-level and word-level features are concatenated before being input into the linear layer for classification.

\subsection{Feature-fusion System}
\label{subsec:linguistic_only}

Swin-BERT is designed to leverage the extracted acoustic information and linguistic information, as is shown in Figure \ref{fig:late_fusion} (c). In our designed feature fusion system, the learned acoustic feature $M_{acoustic} \in \mathcal {R}^{32 \times 1024}$ is the statistical matrix generated by the acoustic feature $X^a \in \mathcal {R}^{1 \times 1024}$ extracted from the final linear layer of the acoustic-based system. While designing the feature fusion system, whether to include the character-level embedding matrix as the extra input for the linguistic-based system is also explored. The results show that when fusing with the learned acoustic features, only word-level transcripts can get a better performance on the two datasets. More discussion are shown in Section \ref{results}.

\section{Experimental Settings}
\label{experimental_setup}

In this section, the information on the ADReSS and ADReSSo datasets, as well as the parameters of the proposed systems, are presented in detail.

\subsection{Dataset Information}

The picture description task is a constrained task that relies less on episodic memory and more on semantic knowledge and retrieval ability \cite{mueller2018connected}. A picture is presented as a prompt in the test, with individuals then being asked to describe what they have seen in this picture. During this process, answers are often recorded to be subsequently used by the neuropsychologist when scoring their tests. The most commonly used picture is that of a line drawing called the “Cookie Theft”, which originates from a test for aphasia \cite{goodglass1972assessment}. The datasets used in this paper are both based upon the cookie theft picture description task and are presented in more detail in Table \ref{tab:dataset_adress} and Table \ref{tab:adresso_data_1}, respectively.

\begin{table}[!htbp]
  \caption{The participant and recording information of the ADReSS dataset.}
  \label{tab:dataset_adress}
    \vspace*{3mm}
  \centering
  \begin{tabular}{p{0.6cm}<{\centering}p{0.8cm}<{\centering}p{0.6cm}<{\centering}p{2.2cm}<{\centering}p{2.2cm}<{\centering}}
    \toprule
    \textbf{Subset} & \textbf{Patient Group} &\textbf{Gender (M:F)}  & \textbf{Age; Means[\#Rec]}  & \textbf{Duration; Means[\#Rec]} \\
    \midrule
    \multirow{2}{*}{Training}&AD & 24:30 & 66.91$\pm$(6.52)[54] & 82.24$\pm$(43.21)[54]\\
    & HC & 24:30& 66.21$\pm$(6.41)[54] & 61.46$\pm$(20.76)[54] \\
    \midrule
    \multirow{2}{*}{Test}&AD  & 11:13 & 66.13$\pm$(7.28)[24] & 90.47$\pm$(51.75)[24]\\
    & HC &11:13& 66.13$\pm$(6.94)[24] & 74.55$\pm$(31.51)[24]\\
    \bottomrule
  \end{tabular}
\end{table}

The ADReSS dataset \cite{LuzHaiderEtAl20ADReSS} shown in Table \ref{tab:dataset_adress} was constructed and shared as part of the Interspeech-2020 ADReSS special session, aiming to provide researchers with a benchmark dataset for linguistic- and acoustic-based dementia detection tasks. The shared dataset provided both acoustic recordings and the corresponding manual transcripts, which is a subset of the DementiaBank dataset and is constructed to consider the balance of gender and age. The ADReSSo dataset \cite{bib:LuzEtAl21ADReSSo} is associated with the Interspeech-2021 ADReSSo special session, and only the acoustic recordings are provided without any of their corresponding manual transcripts. Table \ref{tab:adresso_data_1} summarises the information gathered from the audio recordings and provided in Challenge task 1 (classification task), which includes 122 recordings from the AD group and 115 from the HCs.

\begin{table}[!htbp]
  \caption{The participant and recording information of the ADReSSo data to be used for the binary classification task, \#UNK is used to represent unknown.}
  \label{tab:adresso_data_1}
  \centering
  \begin{tabular}{p{1.0cm}<{\centering}p{1.0cm}<{\centering}p{1.0cm}<{\centering}p{1.0cm}<{\centering}p{2.0cm}<{\centering}}
    \toprule
 \textbf{Subset}&\textbf{Patient Group} &\textbf{Age} &\textbf{Gender}& \textbf{Duration;  Means[\#Rec]} \\
    \midrule
    \multirow{2}{*}{Training}
    & AD  &\#UNK&  \#UNK & 87.61$\pm$(46.31)[87]\\
    &HC &\#UNK& \#UNK & 68.76$\pm$(25.04)[79]\\
    \midrule
    \multirow{2}{*}{Test}
    & AD & \#UNK & \#UNK& 79.42$\pm$(36.26)[35]\\
    &HC & \#UNK & \#UNK & 66.35$\pm$(28.18)[36]\\    
    \bottomrule
  \end{tabular}
\end{table}

In this paper, considering that no manual transcripts are available for the ADReSSo dataset, all the transcripts used are generated by the ASR system. The two datasets have separate test sets, so the systems shown in Section \ref{systems} are trained with the training sets and the results shown in Section \label{results} are based on the test sets.

\subsection{System Parameters}


\subsubsection{Parameters of the acoustic-based System}

The input waveform is first transformed into Mel-spectrograms for the acoustic-based system, with the dimension $F$ set to 64. In the patch embedding stage, the patch window size $P$ is set to 4, allowing for a patch window size of $4 \times 4$. 
Subsequent to this, four Swin Transformer blocks are used, with the number of attention heads in the SW-MSA being 4, 8, 16 and 32 for each Swin Transformer block. The latent size $D$ is set to 96 for all layers. For training purposes, the model is firstly pre-trained on the AudioSet dataset \cite{gemmeke2017audio}, then fine-tuned on ADReSS and ADReSSo datasets. Adam optimiser is used for all experiments with $\beta 1$ = 0.95, $\beta 2$ = 0.999. The initial learning rate is $10^{-4}$. These parameters are selected according to the parameters shown in \cite{chen2022hts}.

\subsubsection{Parameters of the ASR System}
\label{subsubsec:ASR_system}

Wav2vec2.0 is used as the ASR system, which is pre-trained and fine-tuned on 960 hours of Librispeech using 16kHz sampled speech audio. The pre-trained system is then used to generate both word-level and character-level transcripts. No manual transcripts from the ADReSS or ADReSSo datasets are used for fine-tuning the pre-trained wav2vec2.0 system.

\subsubsection{Parameters of the Linguistic-based System}
\label{subsubsec:linguistic_only_system}

The parameter of each layer of the linguistic-based system is shown in Figure \ref{fig:late_fusion}. The input and output dimensions of the first convolution layer are 1024 and 512, respectively, followed by the second convolution layer with 128 output dimensions. \textsl{BERT-base-uncase} is used as the pre-trained model to initialise the structure used for extracting the word-level feature. The output of BERT's last layer is used to represent the word-level features. The batch size is set to 8, and the epoch is equal to 12. The maximum word limit for the word-level transcripts is set to 512, and the maximum character limit for the character-level transcripts is set to 3000, according to the average length of both the word-level and sentence-level transcripts. Each of these parameters is selected according to the quality of the results attained during the training set.

\subsubsection{Parameters of Swin-BERT System}

The acoustic feature matrix $M_{acoustic}$ is padded to [32*1024]. Feature matrices that are shorter than 32 are padded with zero, whilst those longer than 32 are chunked. Swin-BERT is similar to the linguistic-based system, but we used learned acoustic features instead of character-level transcripts as the input. The input and output dimensions of the first convolution layer are 32 and 16, respectively, followed by the second convolution layer with 16 output dimensions. The batch size is set to 8, and the epoch is equal to 12. The maximum word limit of the word-level transcripts is set to 512, and the maximum character limit of the character-level transcripts is set to 1024, which are selected according to the quality of the results attained during the training set.

\section{Results}
\label{results}

This section shows the performance of the acoustic-based, the linguistic-based, and the fusion systems evaluated using the ADReSS and ADReSSo datasets. Results are analysed to demonstrate the efficiency of our proposed methods, and the related results on the two datasets  proposed previously are summarised to compare with our proposed systems.

\subsection{Linguistic-based Results}

\begin{table}[!htbp]
  \caption{Linguistic-based results (\%) with wav2vec2.0 automatic transcripts as the input of BERT. \textsl{extra} refers the character-level embedding matrix.}
  \label{tab:wav2vec_linguistic_res_best}
  \centering
  \begin{tabular}{p{1.0cm}<{\centering}p{1.9cm}<{\centering}p{0.8cm}<{\centering}p{0.8cm}<{\centering}p{0.8cm}<{\centering}p{0.8cm}<{\centering}}
    \toprule
    \textbf{Dataset} & \textbf{System}&\textbf{Accuracy} &\textbf{Precision}  & \textbf{Recall} & \textbf{Fscore}\\
    \midrule
       \multirow{6}{*}{ADReSS}
       & \cite{luz2020alzheimer} & 75.00 & 76.50& 74.50& 74.50\\
    & \cite{koo2020exploiting}&81.25&81.31&81.25& 81.24 \\
    & \cite{cummins2020comparison}&81.30& -& - &81.20 \\
    & \cite{yuan2020disfluencies}&85.40 &\textbf{ 87.00}& 85.40& 85.25 \\
     &\cite{pan2023ps} &82.64&&&82.63\\
    & Ours (w/o extra)& \textbf{85.42} & 85.50& \textbf{85.42} & 85.42\\
      & Ours (w/ extra) & \textbf{85.42}& 86.20& \textbf{85.42} & \textbf{85.47}\\
   
   \midrule
    \multirow{5}{*}{ADReSSo} 
    & \cite{bib:LuzEtAl21ADReSSo} & 77.46& - & - & - \\
    & \cite{syed2021tackling}&84.34& -& - & 84.34\\
    & \cite{pan2021using}& 84.51&84.73&84.57&84.50 \\
    & \cite{pappagariautomatic}& 84.51&\textbf{94.00}&79.00&86.00 \\
    &\cite{pan2023ps} &74.68&&&74.65\\
    & Ours (w/o extra) & 84.51 & 84.88& 84.51 & 84.55\\
    & Ours (w/ extra) & \textbf{85.92} & 88.41& \textbf{85.92} & \textbf{86.07}\\
    
    \bottomrule
  \end{tabular}
\end{table}

To explore how the character-level embedding matrix can affect the performance of the linguistic-based dementia detection system, the results of the systems with and without the character-level embedding matrix are shown in Table \ref{tab:wav2vec_linguistic_res_best}. As shown, by including the character-level embedding matrix as the extra input, the system's performance can be improved on both the ADReSS and ADReSSo datasets. Specifically, for the ADReSSo dataset, the F-score can be improved from 84.55\% to 86.07\%.

\subsection{Acoustic-based Results}

Table \ref{audio_best_results} shows the baseline systems and results obtained using our designed acoustic-based system. As shown, our proposed acoustic-based system can ensure superior performance on both the ADReSS and ADReSSo datasets. For the ADReSSo dataset, the proposed acoustic-based system reached an F-score of 76.04\%. For the ADReSS test set, the result was improved from 73.25\% F-score to 79.14\% F-score after using age and gender as the extra input of our designed system. Compared to previous research, our proposed system can ensure comparable or better performance on the two datasets.

\begin{table}[!htbp]
  \caption{The acoustic-based results (\%) for dementia detection.\textsl{extra} refers to the age and gender information provided by the ADReSS datasets.}
  \label{audio_best_results}
  \centering
  \begin{tabular}{p{1.0cm}<{\centering}p{1.9cm}<{\centering}p{0.8cm}<{\centering}p{0.8cm}<{\centering}p{0.8cm}<{\centering}p{0.8cm}<{\centering}}
    \toprule
    \textbf{Dataset} & \textbf{System}&\textbf{Accuracy} &\textbf{Precision}  & \textbf{Recall} & \textbf{Fscore}\\
    \midrule
    
    \multirow{6}{*}{ADReSS} 
    & \cite{luz2020alzheimer} & 63.50  & 62.50 &  62.00 & 62.00\\

    & \cite{cummins2020comparison}  & 72.90  & - & -&72.60 \\
    & \cite{edwards2020multiscale}& \textbf{79.17}&77.51&\textbf{ 79.17}& 79.13 \\
    &\cite{pan2023ps} &\textbf{81.25} &&& \textbf{81.18}\\
     & Ours (w/o extra) & 73.25 & 73.28 &73.22 & 73.25 \\
    & Ours (w/ extra) & 79.12 & 79.17 & 79.11 &79.14 \\
    \midrule
    \multirow{4}{*}{ADReSSo}
    & \cite{bib:LuzEtAl21ADReSSo} & 64.79  & - & -&-\\
    & \cite{balagopalan2021comparing} &67.60 & 63.60& \textbf{80.00}& 70.80\\
    &  \cite{perez2021influence}&  67.61& 60.00&75.00 & 67.00\\ 
    &\cite{pan2023ps} & \textbf{77.46 }&&&\textbf{77.46}\\
    & Ours (w/o extra) & 76.01&\textbf{76.52} &76.11  &76.04 \\
    \bottomrule
  \end{tabular}
\end{table}

\begin{figure}[!htbp]
\begin{minipage}[b]{0.49\linewidth}
  \centerline{\includegraphics[width=\columnwidth]{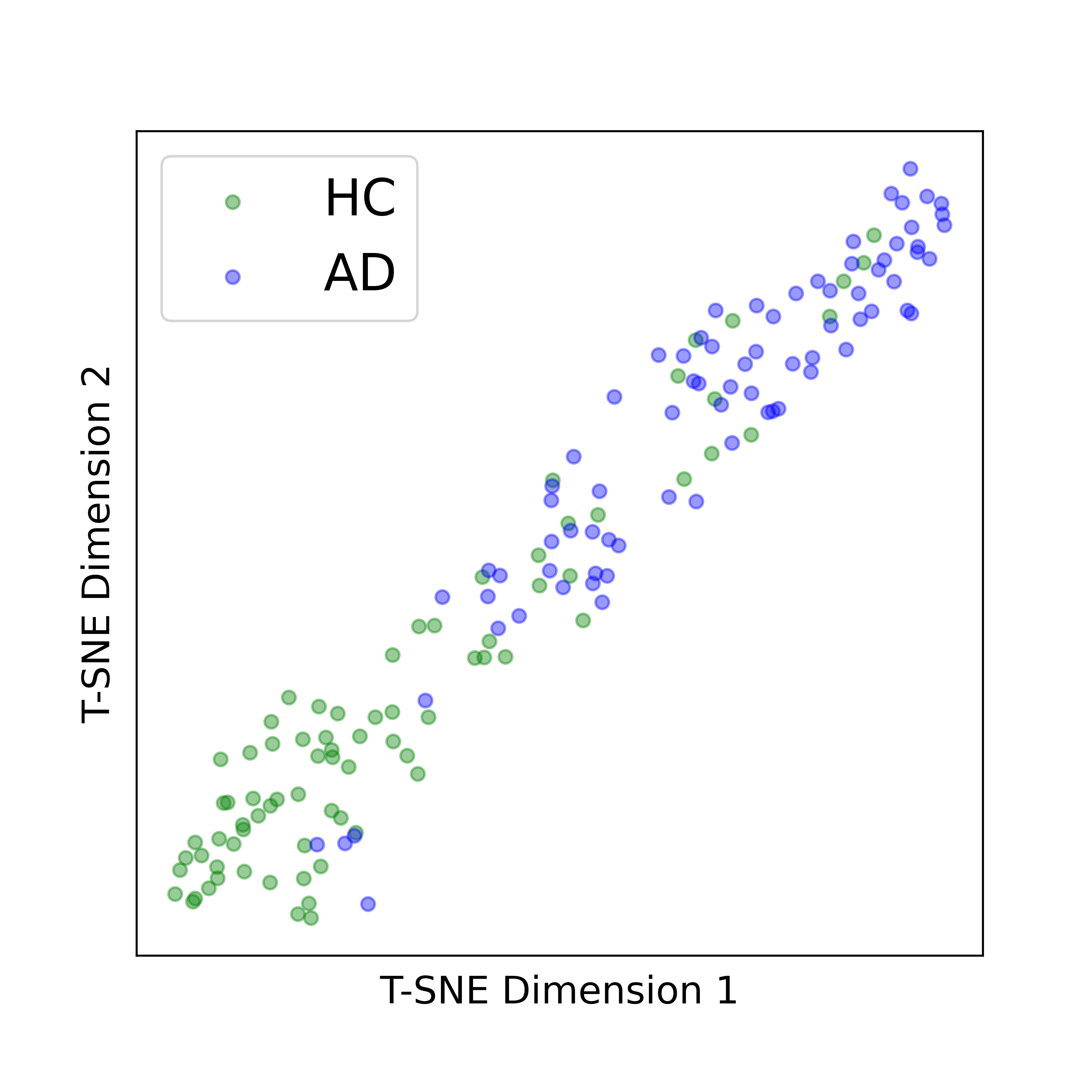}} 
  \label{subfig:wav2vec_acoustic-only}
\end{minipage}
\begin{minipage}[b]{0.49\linewidth}
  \centerline{\includegraphics[width=\columnwidth]{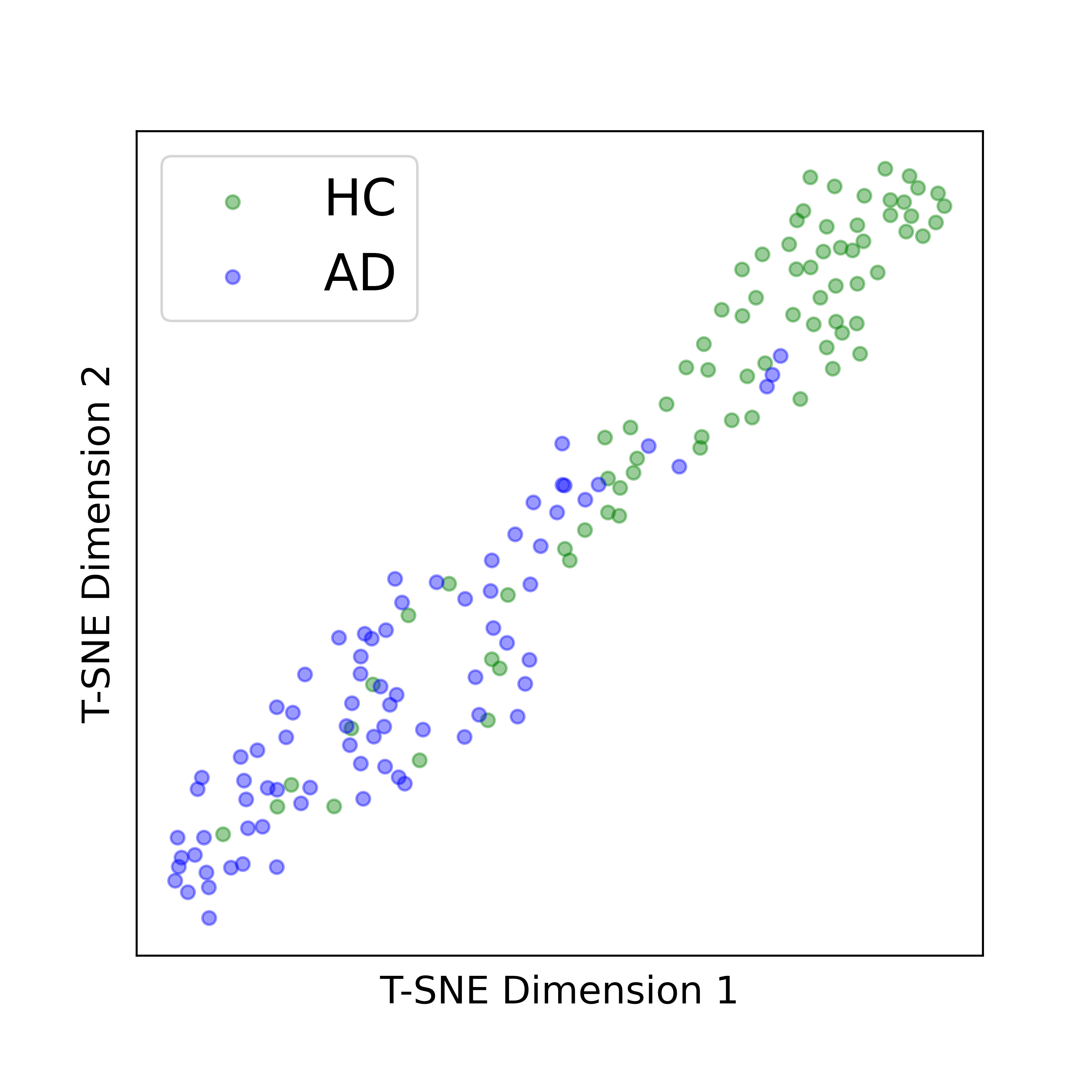}}
  \label{subfig:wav2vec_feature-path_signature}
\end{minipage}

\caption{Embedding visualisation using T-SNE extracted from the ADReSS dataset of (a):  without extra information; (b): with extra information.}
\label{fig:tsne}
\end{figure}

To visualise the effectiveness of using gender and age as the extra information, Figure \ref{fig:tsne} plotted the t-distributed stochastic neighbour embedding (t-SNE) algorithm \cite{van2008visualizing} on features of the ADReSS dataset extracted from the acoustic-based system with and without the age and gender information, respectively. The extracted embedding is projected into two-dimensional space using the t-SNE algorithm. As shown, the data points from the same group in Figure \ref{fig:tsne} (b) are clustered better than the data points in Figure \ref{fig:tsne} (a), which is consistent with the results shown in Table \ref{audio_best_results}.

\subsection{Feature Fusion Results}
\label{subsec:fusion}

Table \ref{tab:wav2vec_fusion_res} shows the feature fusion results based on Swin-BERT on the ADReSS and ADReSSo datasets by using the acoustic features corresponding to the Table \ref{audio_best_results}. Compared to the acoustic-based or linguistic-based results, the results shown in Table \ref{tab:wav2vec_fusion_res} demonstrate that our proposed Swin-BERT system can be improved after doing feature fusion, and ensure state-of-the-art results on the two datasets than previous research. We also explored how age, gender and character-level embedding matrix can affect the performance of the Swin-BERT. The result shows that Swin-BERT can perform better with age and gender information but without the character-level embedding matrix. It is inferred that the learned acoustic features already include rhythm-related information; age and gender information are critical for dementia detection.

\begin{table}[!htbp]
  \caption{Feature-fusion system results (\%) with Swin-BERT.}
  \label{tab:wav2vec_fusion_res}
  \centering
  \begin{tabular}{p{1.0cm}<{\centering}p{1.0cm}<{\centering}p{1.0cm}<{\centering}p{1.0cm}<{\centering}p{1.0cm}<{\centering}p{1.0cm}<{\centering}}
    \toprule
    \textbf{Dataset} & \textbf{System}&\textbf{Accuracy} &\textbf{Precision}  & \textbf{Recall} & \textbf{Fscore}\\
    \midrule
    \multirow{4}{*}{ADReSS} 
    & \cite{koo2020exploiting}& 81.25 &82.80& 81.25& 81.05\\
    & \cite{edwards2020multiscale}& 75.00&75.00&75.00&75.00 \\
    & \cite{cummins2020comparison}&85.20& - & - &85.40\\
    & Ours &\textbf{85.42}& \textbf{87.59}& \textbf{85.42} & \textbf{85.58}\\
    \midrule
    \multirow{4}{*}{ADReSSo} & \cite{bib:LuzEtAl21ADReSSo} & 78.87&78.90& 78.90 & 78.87\\
    & \cite{pappagariautomatic} &84.51& 84.00& 85.50& 84.50 \\
    & \cite{pan2021using}&78.87&79.09& 78.81&78.81 \\
    & Ours &\textbf{87.32}&\textbf{87.36}&\textbf{87.32}&\textbf{87.32} \\
    \bottomrule
  \end{tabular}
\end{table}

\section{Conclusion}\label{conclusion}

This paper proposed a system named Swin-BERT for constructing a speech-based dementia detection system. The proposed system contained two parts: the acoustic-based system and the linguistic-based system. For the acoustic-based system, the SW-MHA is used to extract both local and global features for AD classifications while leveraging the age and gender information. For the linguistic-based system, the character-level embedding matrix is used as the extra input of BERT designed for the word-level transcripts. The proposed feature fusion system named Swin-BERT leverages the local and global information learned from the acoustic-based system together with the linguistic-based system. The result shows that our proposed Swin-BERT can ensure superior results on the ADReSS and ADReSSo datasets.

\newpage
\bibliographystyle{IEEEbib}
\bibliography{ref.bib}

\end{document}